\documentclass[preprint,showkeys,showpacs,aps]{revtex4}

\usepackage{graphicx}
\usepackage{dcolumn}
\usepackage{bm}
\usepackage{threeparttable}
\newcommand{\thickhline}{\noalign{\hrule height 0.8pt}}

\begin{document}


\title{Critical
behavior and universality properties of uniaxial ferromagnetic thin
films in the presence of random magnetic fields}
\author{Yusuf Y\"{u}ksel}
\email{yusuf.yuksel@deu.edu.tr}

\affiliation{Department of Physics, Dokuz Eyl\"{u}l University,
Kaynaklar Campus, TR-35160 Izmir, Turkey}
\date{\today}
\begin{abstract}
Critical phenomena in uniaxial ferromagnetic thin films in the
presence of random magnetic fields have been studied within the
framework of effective field theory. When the type of the random field distribution is bimodal,
the system exhibits tricritical behavior. Furthermore, the critical
value of surface to bulk ratio of exchange interactions at which the
transition temperature becomes independent of film thickness is
insensitive to the presence of disorder whether the distribution is
bimodal or trimodal. Regarding the universality
properties, neither $p$, nor $h$ variations in the system can affect
the value of the shift exponent $\lambda$. In this regard, it can be
concluded that pure ferromagnetic thin films are in the same
universality class with those under the influence of random discrete
magnetic fields.
\end{abstract}

\maketitle

\section{Introduction}\label{intro}
The surface magnetism which was proposed approximately four decades
ago from present by Mills \cite{mills0,mills1} is still one of the
most actively studied research fields in statistical mechanics of
phase transitions and critical phenomena
\cite{kaneyoshi1,pleimling0}. The reason is due to the fact that in
the presence of free surfaces, magnetic properties of thin films
drastically differ from those of bulk materials. In other words, due
to their reduced coordination number, the surface atoms have less
translational symmetry and the exchange interaction between two
adjacent surface atoms is different than that of the inner atoms
\cite{kaneyoshi2}. As a consequence of these facts, the surface may
exhibit an ordered phase even if the bulk itself is disordered. This
phenomenon has already been experimentally observed
\cite{ran,polak,tang}. In this context, an extraordinary case is
defined as the transition at which the surface becomes disordered at
a particular temperature $T_{c}^{s}$ which is larger than the bulk
transition temperature $T_{c}^{b}$, on the contrary, in an ordinary
case, bulk region orders when the surface is disordered.

It is theoretically predicted that there exists a critical value of
surface to bulk ratio of exchange interactions $R_{c}$ above which
the surface effects are dominant and the transition temperature of
the entire film is determined by the surface magnetization whereas
below $R_{c}$, the transition characteristics of the film are
governed by the bulk magnetization. The critical value $R_{c}$
itself is called as the special point, and the numerical value of
this point has been examined within various theoretical techniques
based on several extensions of an Ising type spin Hamiltonian
\cite{aguilera,binder,kaneyoshi3,kaneyoshi4,burkhardt,landau,sarmento0,neto,tucker}.
Among these works, within the framework of effective field theory
(EFT), Sarmento and Tucker \cite{sarmento0} clarified that a
transverse field in the surface layer causes the critical value of
the surface exchange enhancement $R_{c}$ to move to a higher value
whereas the presence of a bulk transverse field causes $R_{c}$ to
decrease to a lower value. In addition, using extensive Monte Carlo
(MC) simulations, the effect of surface exchange enhancement on
ultra thin spin-1 films has been studied by Tucker \cite{tucker},
and it was concluded that the $R_{c}$ value is spin dependent.
Moreover, in a very recent work, the problem has also been handled
for the systems in the presence of quenched dilute and trimodal
random crystal fields within the framework of EFT, and it has been
reported that the $R_{c}$ value can be modified in the presence of
crystal field disorder \cite{yuksel0}.

In a recent paper, regarding the presence of random magnetic fields,
the effect of Gaussian random longitudinal magnetic field
distribution on the phase diagrams and magnetization behavior of the
transverse Ising thin film have been investigated, and it has been
shown that phase diagrams of the model exhibit only second-order
phase transition properties, and changing the width $\sigma$ of the
random field distribution makes no significant change in $R_{c}$
\cite{akinci}.

On the other hand, theoretical and experimental investigations are
also focused on the finite size shift of the critical temperature of
the film as a function of its thickness which is characterized by a
shift exponent $\lambda$. A number of experimental studies have been
devoted to determine the value of $\lambda$ for various thin film
samples, and it has been concluded that the shift exponent extends
from $\sim1$ to $3.15$ \cite{farle,elmers,henkel,ballentine}. Since
the exponent $\lambda$ is directly related on the bulk correlation
length exponent as $\nu_{b}=1/\lambda$ \cite{barber0}, within the
accuracy of Ising-type models, it can be mentioned that a sample of
thin film for which the exponent $\lambda$ is close to unity
exhibits a two dimensional character whereas as the value of the
exponent becomes larger than unity then the system shows a three
dimensional character. Theoretically, the exponent $\lambda$ has
been extracted for some certain models with a wide variety of
techniques. For instance, using the high temperature series
expansion (HTSE) method, it has been shown that the estimated value
of $\lambda$ for ferromagnetic Ising \cite{allan,capehart} and
Heisenberg \cite{ritchie} thin films severely depends on whether a
periodic or free boundary condition was considered in the surface.
This result has also been verified within the resolution of MC
simulations \cite{binder2,kitatani,takamoto,laosiritaworn}.
Moreover, the value of the extracted exponent is also very sensitive
to the lattice geometry \cite{masrour}.

Under certain circumstances, universal behavior of a thin film
system may experience a dimensional crossover. Such a phenomenon has
been experimentally observed as the film thickness is varied in
ultra-thin $\mathrm{Ni}(1 1 1)$ films on $\mathrm{W}(1 1 0)$
\cite{li}, and epitaxial thin films of Co, Ni, and their alloys
grown on $\mathrm{Cu}(1 0 0)$ and $\mathrm{Cu}(1 1 1)$ \cite{huang}.
Previous MC simulations \cite{binder} also predict that the exponent
$\lambda$ may vary continuously with surface exchange $J_{s}$ in the
range $R<R_{c}$ which also indicates the occurrence of a dimensional
crossover between the surface value and bulk value.

In our previous work \cite{yuksel1}, with using EFT, we have studied
the universal behavior and critical phenomena in a ferromagnetic
thin film described by a spin-1 Blume-Capel Hamiltonian
\cite{yuksel1} and we have found that the critical value of surface
to bulk ratio of exchange interactions $R_{c}$ strictly depends on
the crystal field interactions. On the other hand, our numerical
results yield that in terms of the exponent $\lambda$, a
ferromagnetic spin-1/2 thin film is in the same universality class
with its spin-1 counterpart.

Based on these circumstances, in the present paper, we extended our
study, and we investigated the phase transition properties of
ferromagnetic uniaxial Ising thin films in the presence of random
magnetic fields. We also extracted the shift exponent for the
present model, and discussed the universality properties of the
system in the presence of random fields by examining the variation
of the exponent $\lambda$ as a function of random field parameters.

The organization of the paper is as follows: In Section \ref{formulation} we briefly
present the formulations. The results and discussions are presented
in Section \ref{results}, and finally Section \ref{conclude} contains our conclusions.

\section{Model and Formulation}\label{formulation}
We consider a ferromagnetic thin film with thickness $L$ in the presence of random magnetic fields described by
the following Hamiltonian
\begin{equation}\label{eq1}
\mathcal{H}=-\sum_{<ij>}J_{ij}S_{i}S_{j}-\sum_{i}h_{i}S_{i},
\end{equation}
where $J_{ij}=J_{s}$ if the lattice sites $i$ and $j$ belong to one of the two surfaces of the film, otherwise we have $J_{ij}=J_{b}$ where $J_{s}$ and $J_{b}$
denote the ferromagnetic surface and bulk exchange interactions, respectively. The first term in Eq. (\ref{eq1}) is a summation over the nearest-neighbor
spins with $S_{i}=\pm1$ and the second term represents the Zeeman energy originating from spatially random magnetic fields on the lattice which are
distributed according to a given probability distribution function. The present study deals with a trimodal distribution which is defined as
\begin{equation}\label{eq2}
P(h_{i})=p\delta(h_{i})+\left(\frac{1-p}{2}\right)[\delta(h_{i}-h_{0})+\delta(h_{i}+h_{0})].
\end{equation}
According to Eq. (\ref{eq2}), we have a pure system for $p=1.0$,
whereas as $p$ approaches to zero, the form of the random field
distribution becomes a bimodal-type where half of the lattice sites
are subject to a magnetic field $-h_{0}$ and the remaining lattice
sites have a field $h_{0}$. Then $p$ can be regarded as an
adjustable parameter which controls the amount of disorder in the
system.

The magnetizations $m_{i}$ ($i=1,...,L$) perpendicular to the surface of the film corresponding to $L$ parallel distinct layers can be obtained by conventional
EFT formulation based on differential operator technique and decoupling approximation (DA) \cite{kaneyoshi0}
\begin{eqnarray}\label{eq3}
\nonumber
m_{1}&=&[A_{1}+m_{1}B_{1}]^{z}[A_{2}+m_{2}B_{2}]F(x)|_{x=0},\\
\nonumber
m_{l}&=&[A_{2}+m_{l}B_{2}]^{z}[A_{2}+m_{l-1}B_{2}][A_{2}+m_{l+1}B_{2}]F(x)|_{x=0},\\
m_{L}&=&[A_{1}+m_{L}B_{1}]^{z}[A_{2}+m_{L-1}B_{2}]F(x)|_{x=0},
\end{eqnarray}
where $2\leq l \leq L-1$, and the coefficients $A_{i}$ and $B_{i}$ are defined as $A_{1}=\cosh(J_{s}\nabla)$, $A_{2}=\cosh(J_{b}\nabla)$,
$B_{1}=\sinh(J_{s}\nabla)$ and $B_{2}=\sinh(J_{b}\nabla)$. In the present work, we will focus on the ferromagnetic films in a
simple cubic lattice structure, i.e. $z=4$ where $z$ is the intra-layer coordination number. The function $F(x)$ in Eq. (\ref{eq3}) is
then given by
\begin{equation}\label{eq4}
F(x)=\int dh_{i}P(h_{i})\tanh[\beta (x+h_{i})],
\end{equation}
where $\beta$ is the inverse of the reduced temperature.

Using the Binomial expansion
\begin{equation}\label{eq5}
(x+y)^{n}=\sum_{i=0}^{n}
\left(\begin{tabular}{c }
  n  \\
  i  \\
\end{tabular}\right)x^{n-i}y^{i},
\end{equation}
in Eq. (\ref{eq3}) we get
\begin{eqnarray}\label{eq6}
\nonumber
m_{1}&=&\sum_{i=0}^{z}\sum_{j=0}^{1}K_{1}(i,j)m_{1}^{i}m_{2}^{j},\\
\nonumber
m_{l}&=&\sum_{i=0}^{z}\sum_{j=0}^{1}\sum_{k=0}^{1}K_{2}(i,j,k)m_{l}^{i}m_{l-1}^{j}m_{l+1}^{k},\\
m_{L}&=&\sum_{i=0}^{z}\sum_{j=0}^{1}K_{1}(i,j)m_{L}^{i}m_{L-1}^{j},
\end{eqnarray}
where
\begin{eqnarray}\label{eq7}
\nonumber
K_{1}(i,j)&=&\left(\begin{tabular}{c}
  z  \\
  i  \\
\end{tabular}\right)
A_{1}^{z-i}A_{2}^{1-j}B_{1}^{i}B_{2}^{j}F(x)|_{x=0},\\
K_{2}(i,j,k)&=&
\left(\begin{tabular}{c}
  z  \\
  i  \\
\end{tabular}\right)
A_{2}^{z+2-i-j-k}B_{2}^{i+j+k}F(x)|_{x=0}.
\end{eqnarray}
Consequently, applying the Binomial expansion for the coefficients $A_{i}$ and $B_{i}$ $(i=1,2)$ in Eq. (\ref{eq7}) yields
\begin{eqnarray}\label{eq8}
\nonumber
K_{1}(i,j)&=&
\left(\begin{tabular}{c}
  z  \\
  i  \\
\end{tabular}\right)
2^{-(z+1)}
\sum_{x=0}^{z-i}\sum_{y=0}^{1-j}\sum_{r=0}^{i}\sum_{t=0}^{j}(-1)^{t+r}
\left(\begin{tabular}{c }
  z-i  \\
  x  \\
\end{tabular}\right)
\left(\begin{tabular}{c }
  1-j  \\
  y  \\
\end{tabular}\right)
\left(\begin{tabular}{c }
  i  \\
  r  \\
\end{tabular}\right)
\left(\begin{tabular}{c }
  j  \\
  t  \\
\end{tabular}\right)
\\
\nonumber
&&\times\exp[(z-2r-2x)J_{s}\nabla]\exp[(1-2y-2t)J_{b}\nabla]F(x)|_{x=0},\\
\nonumber
K_2(i,j,k)&=&
\left(\begin{tabular}{c}
  z  \\
  i  \\
\end{tabular}\right)
2^{-(z+2-2k)}\sum_{x=0}^{(z+2-i-j-k)}\sum_{y=0}^{(i+j-k)}(-1)^{y}
\left(\begin{tabular}{c }
  z+2-i-j-k  \\
  x  \\
\end{tabular}\right)
\left(\begin{tabular}{c }
  i+j-k  \\
  y  \\
\end{tabular}\right)\\
&&
\times\exp[(z+2-2k-2x-2y)J_{b}\nabla]F(x)|_{x=0}.
\end{eqnarray}
With the help of the relation $\exp(\alpha\nabla)F(x)=f(x+\alpha)$
for an arbitrary $\alpha$, the coefficients $K_{1}$ and $K_{2}$ can
be numerically evaluated. Hence, by inserting Eq. (\ref{eq8}) in Eq.
(\ref{eq6}) we obtain a system of coupled non-linear equations which
contains $L$ unknowns which are nothing but just the layer
magnetizations of the film. The longitudinal magnetization $m_{i}$
of each layer can be obtained from numerical solution of Eq.
(\ref{eq6}). Then the bulk and surface magnetizations of a
ferromagnetic thin film can be defined as
\begin{equation}\label{eq9}
m_{b}=\frac{1}{L-2}\sum_{i=2}^{L-1}m_{i}, \quad
m_{s}=\frac{1}{2}(m_{1}+m_{L}).
\end{equation}

Since, the magnetization of the entire system is close to zero in the vicinity of the second order phase transition, the transition
temperature can be obtained by linearizing Eq. (\ref{eq6}), i.e.
\begin{eqnarray}\label{eq10}
\nonumber
m_{1}&=&K_{1}(1,0)m_{1}+K_{1}(0,1)m_{2},\\
\nonumber
m_{l}&=&K_{2}(1,0,0)m_{l}+K_{2}(0,1,0)m_{l-1}+K_{2}(0,0,1)m_{l+1},\\
m_{L}&=&K_{1}(1,0)m_{L}+K_{1}(0,1)m_{L-1}.
\end{eqnarray}

Critical temperature as a function of the system parameters can be determined from $\mathrm{det}(A)=0$ where $A$ is the coefficients matrix of the set of $L$
linear equations in Eq. (\ref{eq10}). We note that the determination of the transition temperature should be treated carefully since as it was previously stated
in Ref. \cite{sarmento0}, from the many formal solutions of $\mathrm{det}(A)=0$, we have to choose the one corresponding to the highest possible transition temperature.
At this point, we should also note that transition temperature $T_{c}(\infty)$ of the bulk ferromagnetic system in the presence of random fields can be
evaluated by solving the following equation \cite{kaneyoshi0}
\begin{equation}\label{eq11}
1=q\cosh^{q-1}(J_{b}\nabla)\sinh(J_{b}\nabla)F(x)|_{x=0},
\end{equation}
where $J_{b}$ is the ferromagnetic bulk exchange interaction, and $q=6$ corresponding to simple cubic lattice structure.

According to the finite-size scaling theory \cite{barber0}, the deviation of the thickness dependent critical temperature $T_{c}(L)$ of a thin ferromagnetic film
from the bulk critical temperature $T_{c}(\infty)$ can be measured for sufficiently thicker films in terms of a scaling relation:
\begin{equation}\label{eq12}
\varepsilon=1-T_{c}(L)/T_{c}(\infty)\propto L^{-\lambda},
\end{equation}
where $\lambda$ is called the shift exponent which is related to the correlation length exponent of the bulk system as
$\lambda=1/\nu_{b}$. The exponent $\lambda$ can be extracted from numerical data by plotting $\varepsilon$ versus $L$ curves for sufficiently thick
films in a log-log scale then fitting the resultant curve using the standard linear regression method.

\section{Results and Discussion}\label{results}
In this section, we will discuss how the presence of random fields
affects the critical and universal behavior of ferromagnetic
uniaxial thin films. Before proceeding, let us note that the
adjustable system parameters such as temperature, magnetic field,
and surface exchange couplings are measured in terms of the bulk
exchange interactions. Hence, we use the following dimensionless
variables in our calculations: The temperature is defined as
$k_{B}T/J_{b}$, magnetic field is scaled as $h=h_{0}/J_{b}$, and the
surface exchange interactions are defined as $R=J_{s}/J_{b}$.

When the variation of the transition temperature of a ferromagnetic
thin film is plotted as a function of $R$ for different film
thickness $L$, at a particular $R=R_{c}$ value, the transition
temperature becomes independent of film thickness. For $R<R_{c}$
(ordinary case), bulk is magnetically dominant against surface
whereas for $R>R_{c}$ (extraordinary case), ferromagnetism of
surface is enhanced in comparison with the bulk magnetization. In
the ordinary case, thicker films exhibit greater transition
temperatures while in the extraordinary case, one has just the
opposite scenario.

After this short summary, let us divide our study into two parts:
First, we will focus our attention on the effect of random fields
which are distributed according to a symmetric bimodal distribution
which can be achieved by putting $p=0$ in Eq. (\ref{eq2}), then we
will generalize the study for trimodal random fields where we can
use arbitrary $p$ values within the range $0\leq p\leq1$.

\subsection{Bimodal distribution}
\begin{figure}[!h]
\center
\includegraphics[width=8.0cm]{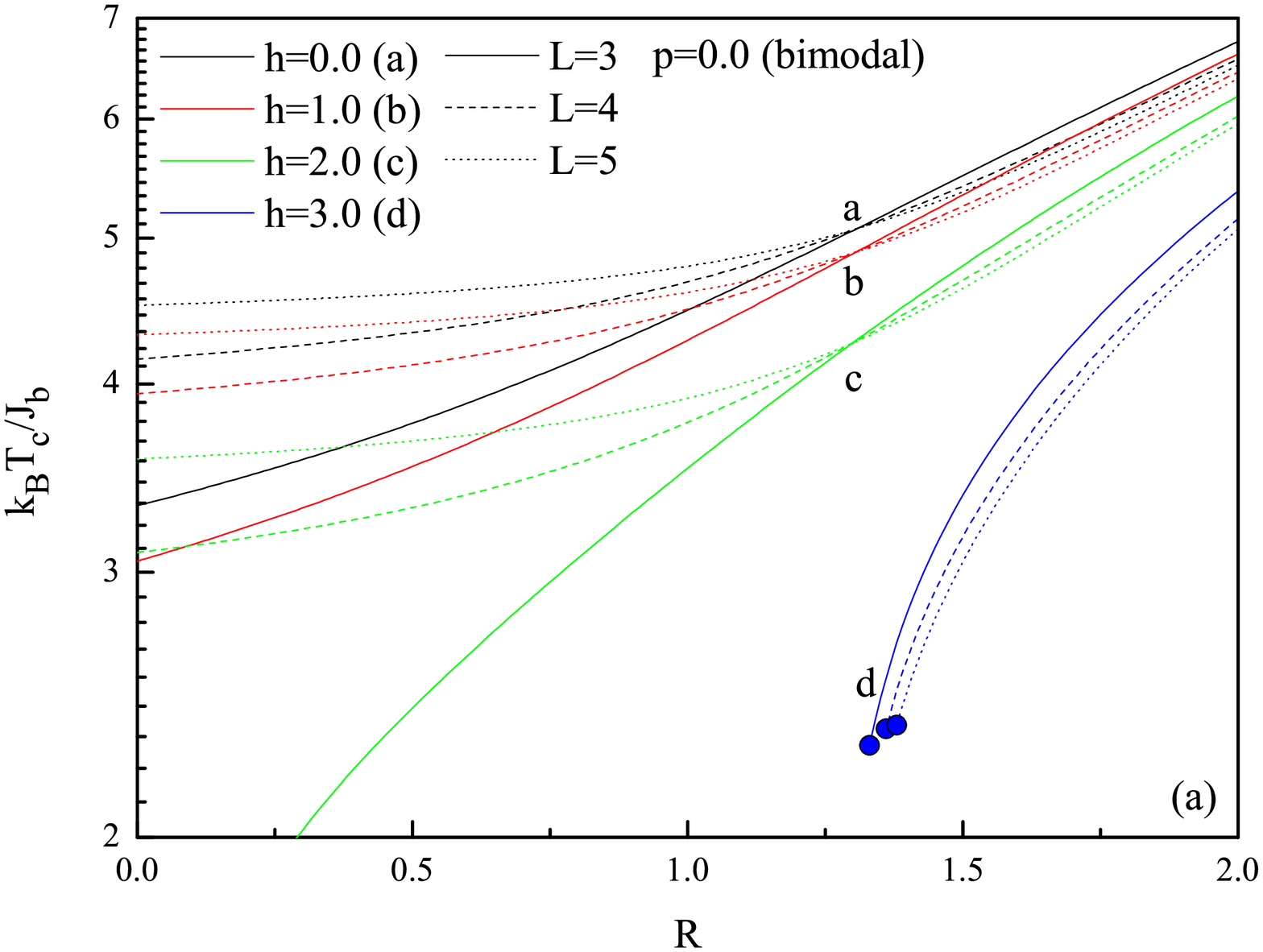}
\includegraphics[width=7.75cm]{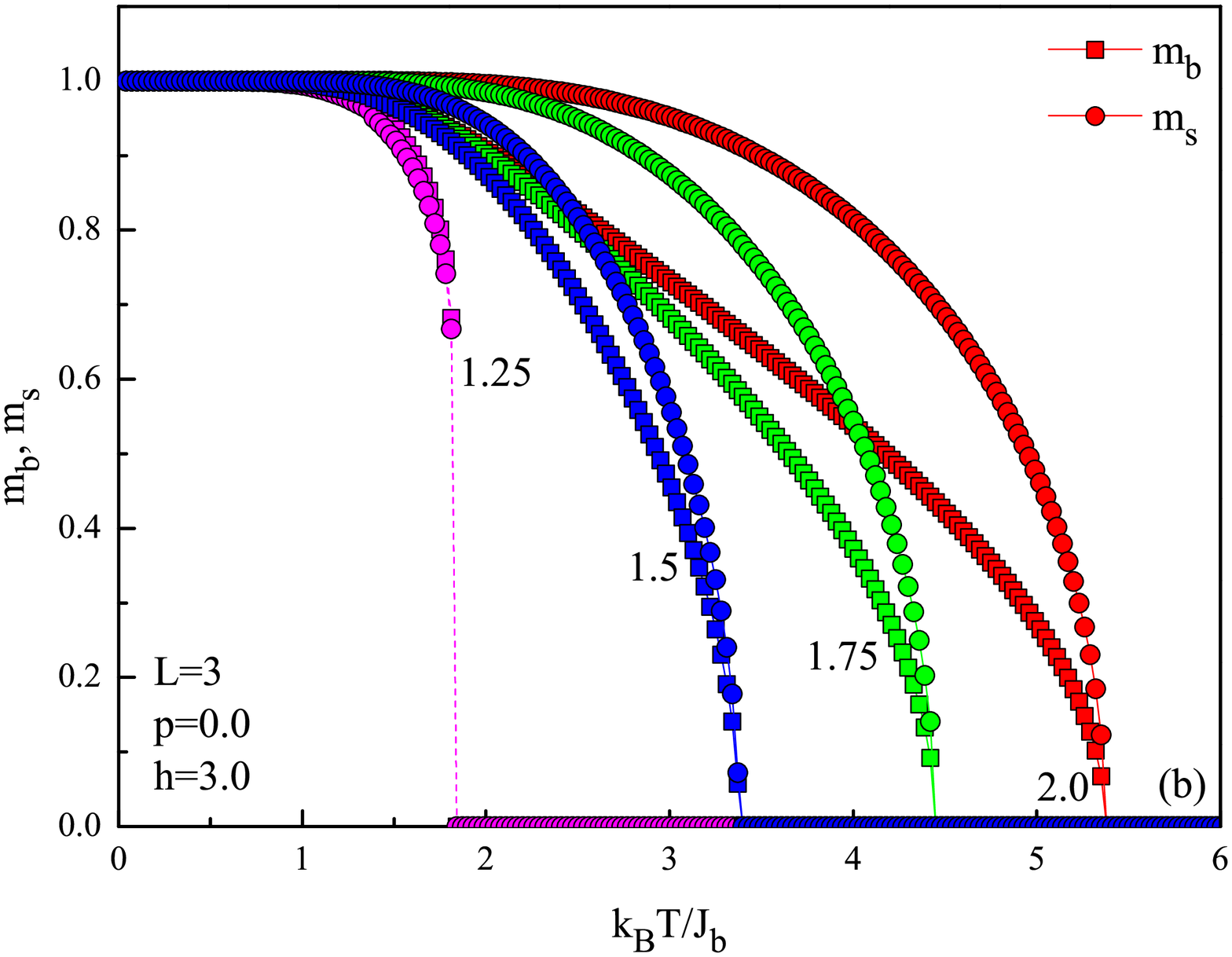}\\
\caption{(a) Phase diagrams of the system in a
$(k_{B}T_{c}/J_{b}-R)$ plane for various film thickness $L$
corresponding to bimodal random field distribution with $h=0.0$,
$1.0$, $2.0$ and $3.0$. Solid circles denote the tricritical points.
(b) Thermal variation of bulk (filled squares) and surface (filled circles) magnetizations. The
numbers accompanying the curves denote the value of the reduced
surface exchange interaction $R$.} \label{fig1}\end{figure}

From Eq. (\ref{eq2}), we can see that a bimodal distribution of
random fields is governed by only one parameter which is the
external field value $h$. In Fig. \ref{fig1}a, in order to see the
effect of external field $h$ on the critical behavior of the system,
we plot the phase diagrams in a $(k_{B}T_{c}/J_{b}-R)$ plane with
different film thickness $L$ and for some selected $h$ values. As
seen in this figure, transition temperature of the films with the
same thickness $L$ decreases with increasing disorder (i.e. as $h$
increases). However, the location of $R_{c}$ does not exhibit any
significant variation with increasing $h$. In other words,
$R_{c}=1.3068$ value obtained for pure films \cite{kaneyoshi4} is
independent of the magnetic field disorder. However, for
sufficiently high $h$ values such as $h=3.0$, randomness effects
become prominent. Consequently, the curves with different $L$ values
do not intersect each other, and the system reaches to its
ferromagnetic phase stability limit by exhibiting a tricritical
behavior.

The linearized equations given in Eq. (\ref{eq10}) are
inapplicable for the first order transitions. However, these
unstable solutions can be located by examining the discontinuous
jumps in thermal dependence of magnetization curves (c.f. Fig.
\ref{fig1}b). We believe that the tricritical behavior observed in
this case originates from the type of the random field distribution.
Namely, the existence of tricritical point may be explained by the following 
physical reason: When the distribution of random magnetic fields is
bimodal, there exist two equal positive and negative large fields. If these large magnetic
fields are dominant against the ferromagnetic exchange interactions such as 
$h=3.0>>R$, in this case, spin-up and spin-down states are canceled out, and the overall 
system behaves like spin-$0$. This concept is very similar to that observed in 
the spin-1 Ising system, when the crystal field $D$
takes a large negative critical value \cite{kaneyoshi0}.  Besides, in the 
presence of a continuous probability
distribution such as a Gaussian, the system has been found to
exhibit only second order phase transitions \cite{akinci}.
Moreover, similar arguments have also been reported for bulk
systems for which different random field distributions lead to
different phase diagrams \cite{yuksel2}.

\begin{figure}[!h]
\center
\includegraphics[width=8.0cm]{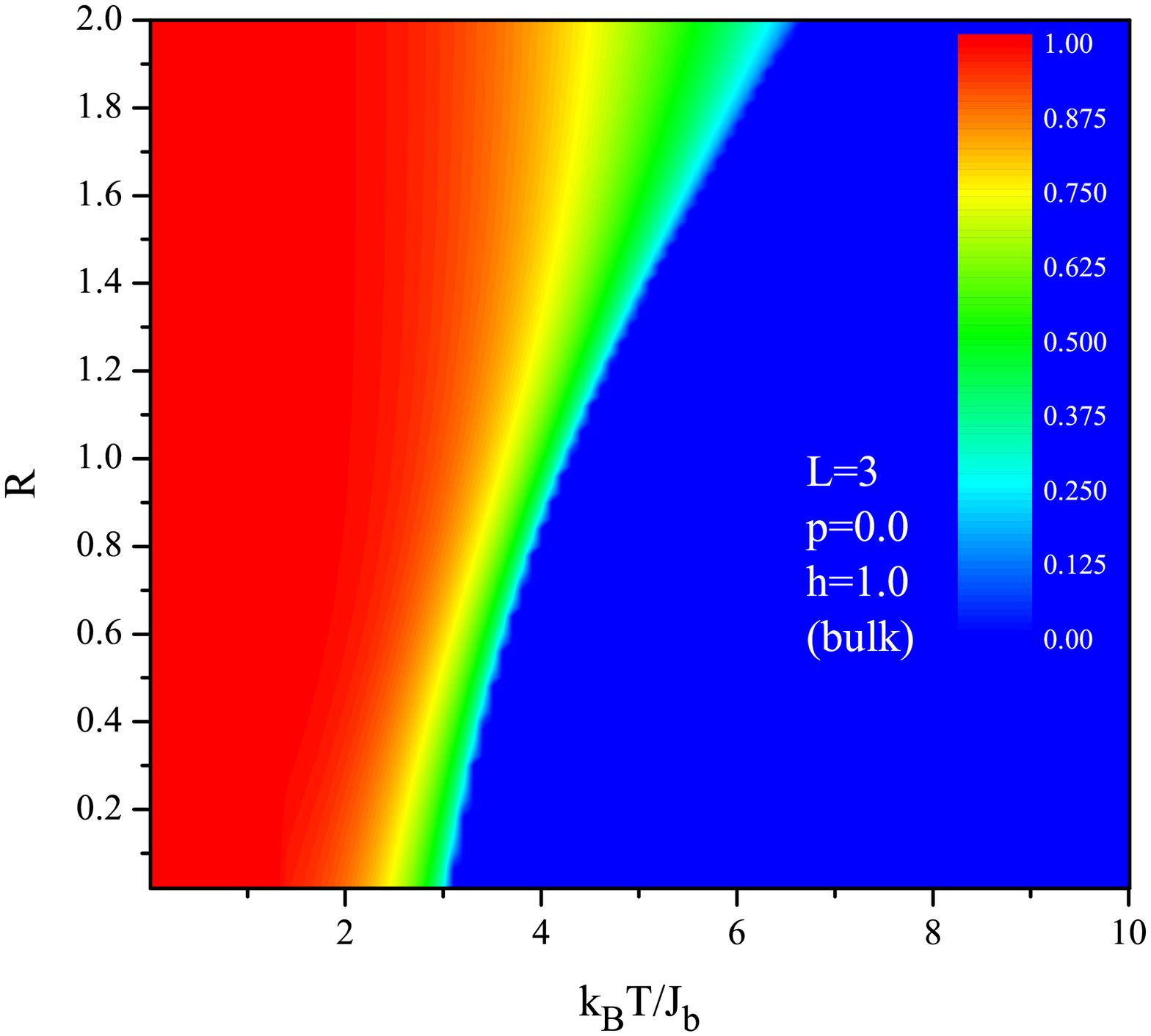}
\includegraphics[width=8.0cm]{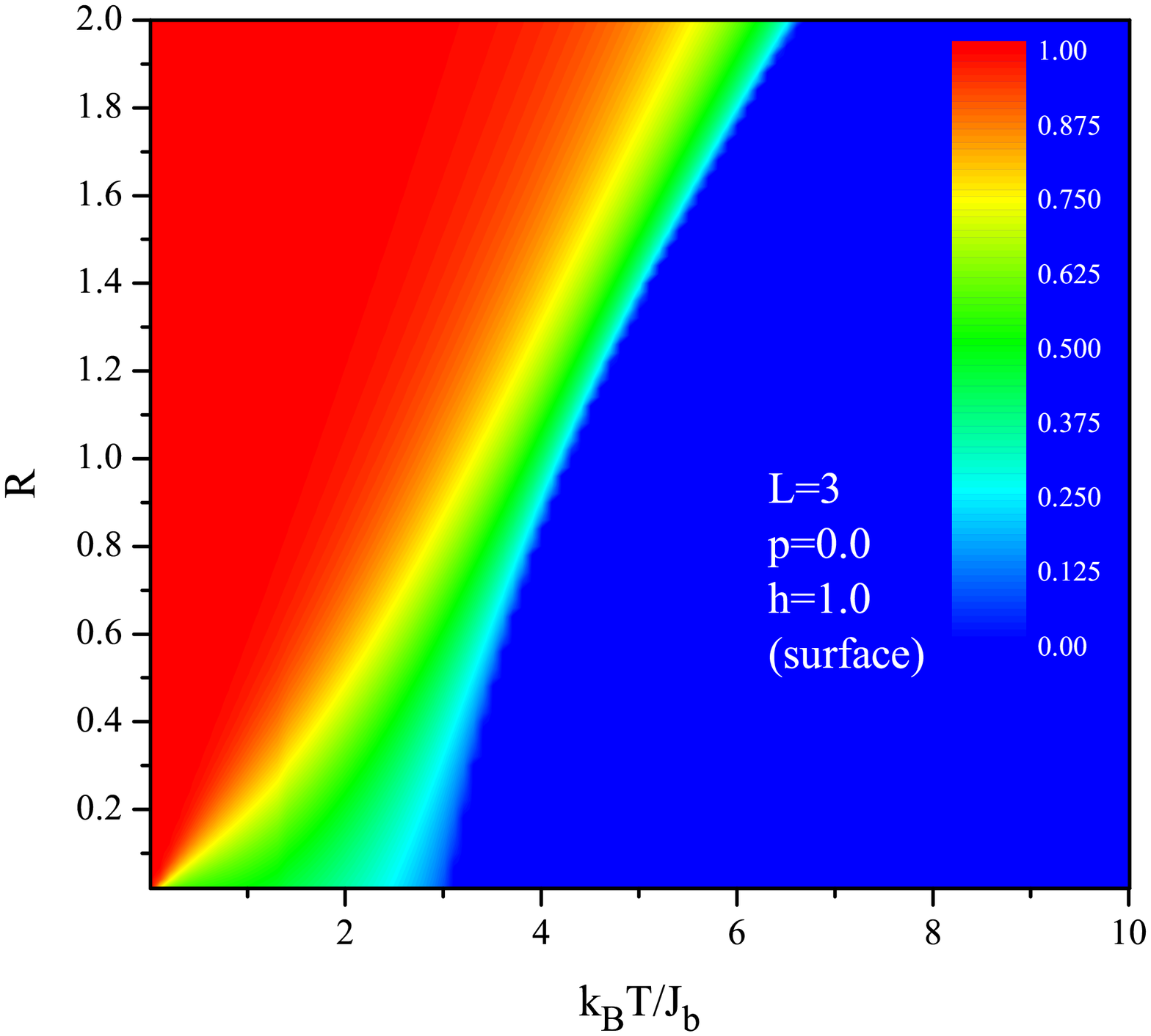}\\
\caption{Contour plot projections of bulk (left panel) and surface
(right panel) magnetizations of a thin ferromagnetic film with $L=3$
in a $(R-k_{B}T/J_{b})$ plane corresponding to the phase diagram
with $h=1.0$ curve depicted in Fig. \ref{fig1}a.} \label{fig2}
\end{figure}

Bulk and surface magnetization profiles as functions of the
temperature and reduced surface exchange interactions are
illustrated in Fig. \ref{fig2} for $L=3$ and $h=1.0$ corresponding
to the curves depicted in Fig. \ref{fig1}a. For weak $R$ values,
ferromagnetism in the bulk region of the film is enhanced against
the surface. However, as $R$ has a value far above $R_{c}$,
ferromagnetic order at the surface region of the film becomes
dominant against the bulk region.

\begin{figure}[!h]
\center
\includegraphics[width=10.0cm]{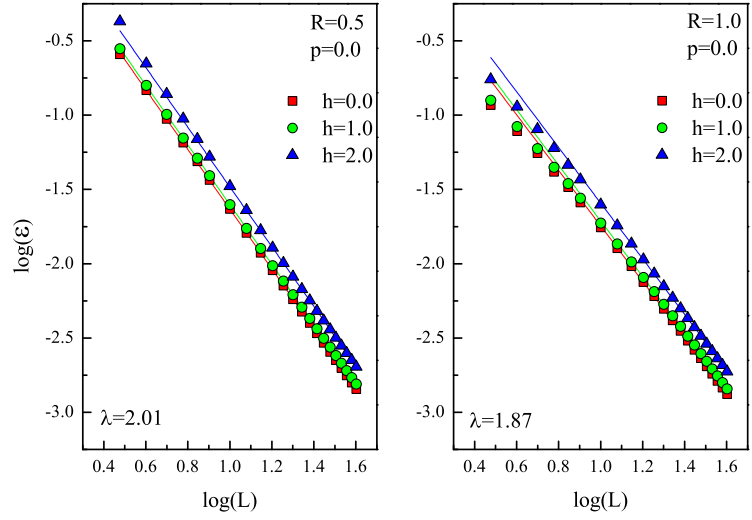}\\
\caption{Variation of the shift exponent $\lambda$ for weak (left
panel) and moderate (right panel) surface couplings corresponding to
bimodal distribution of random fields with some selected values of
magnetic field $h$.} \label{fig3}
\end{figure}

Now, let us discuss the universality properties of the system in the
presence of bimodal random fields. In our previous work
\cite{yuksel1}, we have found that ferromagnetic thin films may
exhibit dimensional crossover as $R$ changes. Namely, as we stated
before, as the value of the shift exponent $\lambda$ approaches to
unity then the system has a two dimensional behavior which is a direct consequence
of the surface exchange enhancement whereas as the bulk region becomes
dominant against the surface, $\lambda$ becomes larger than unity, and
the system shows three
dimensional character. In this context, our previous work yields
that for $R=0.5$, we obtain $\lambda=2.01$, for $R=1.0$ the
exponent value is $\lambda=1.87$, and for $R=1.3$, we have
$\lambda=1.08$ \cite{yuksel1}. These exponent values were extracted
for pure system. However, as shown in Fig. \ref{fig3}, the exponent
$\lambda$ is independent of the randomness. We note that in order to
precisely cover the critical region, the obtained data have been
fitted for those providing the condition
$0.01\leq\varepsilon\leq0.1$ \cite{binder2} which generally requires
to consider the transition temperatures of the films with $L\geq10$
in fitting procedure. Extracted exponent values are summarized in
Table-\ref{table1} for a variety of set of system parameters.

\subsection{Trimodal distribution}
\begin{figure}[!h]
\center
\includegraphics[width=8.0cm]{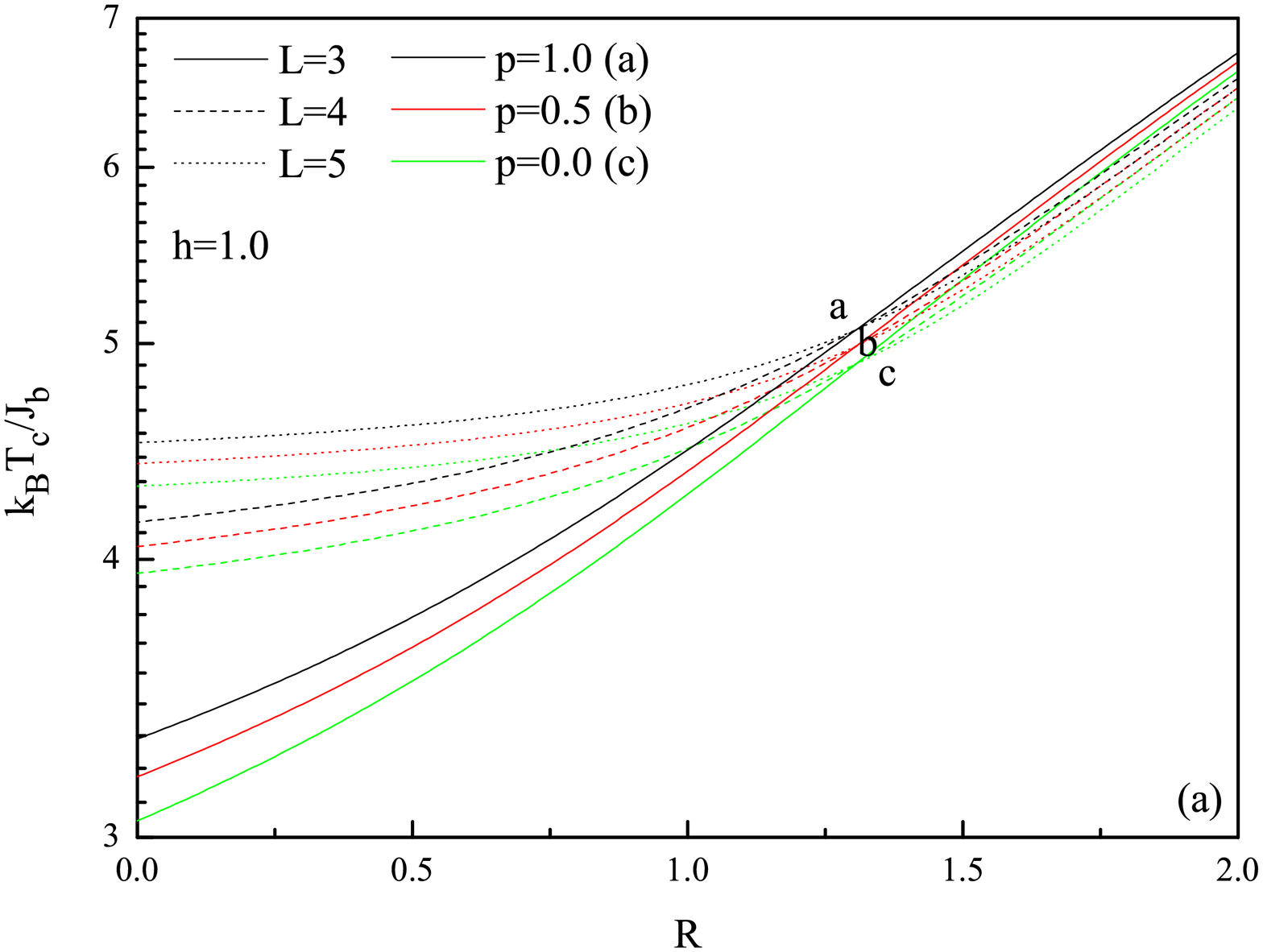}\\
\includegraphics[width=8.0cm]{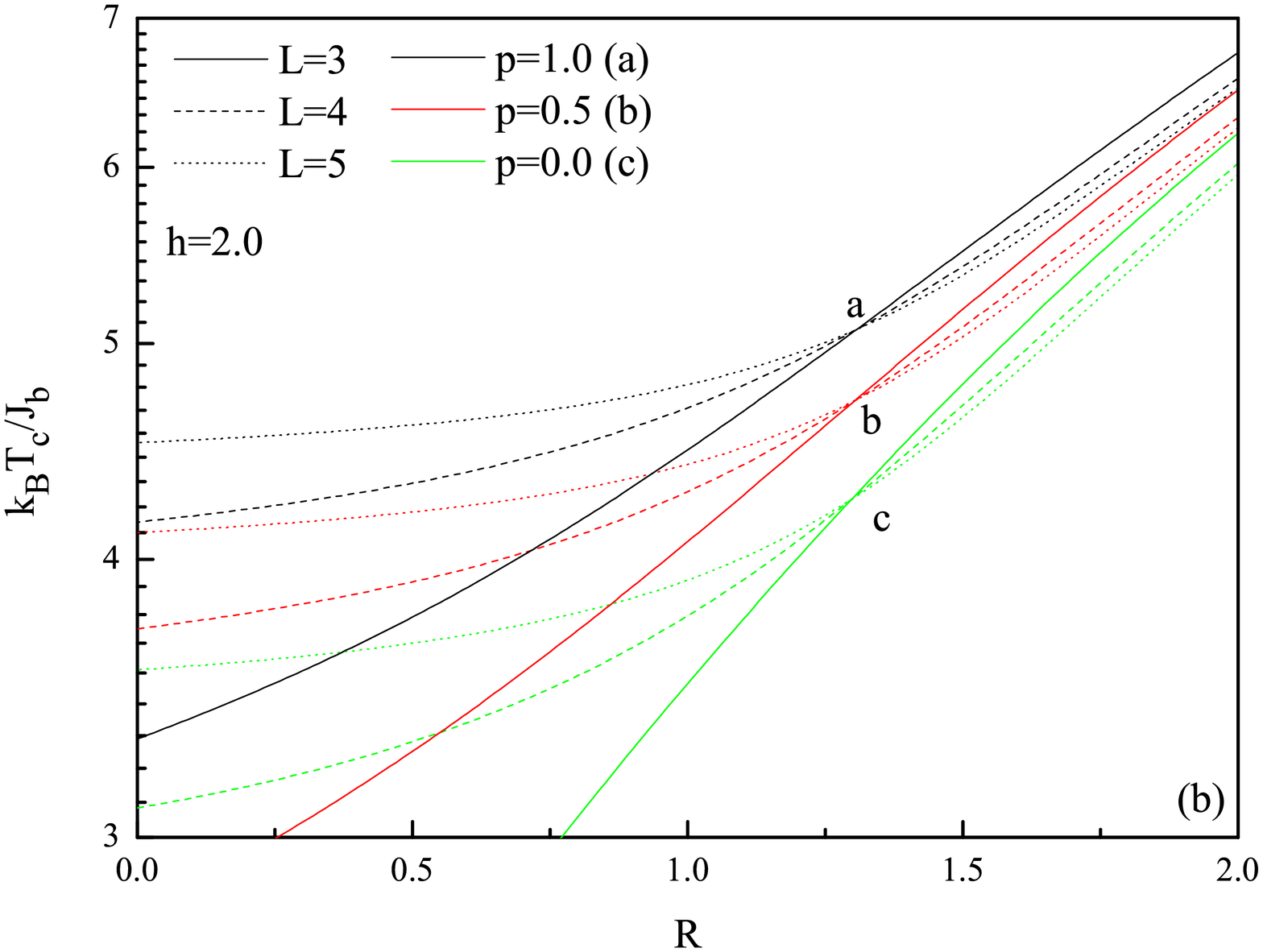}\\
\includegraphics[width=8.0cm]{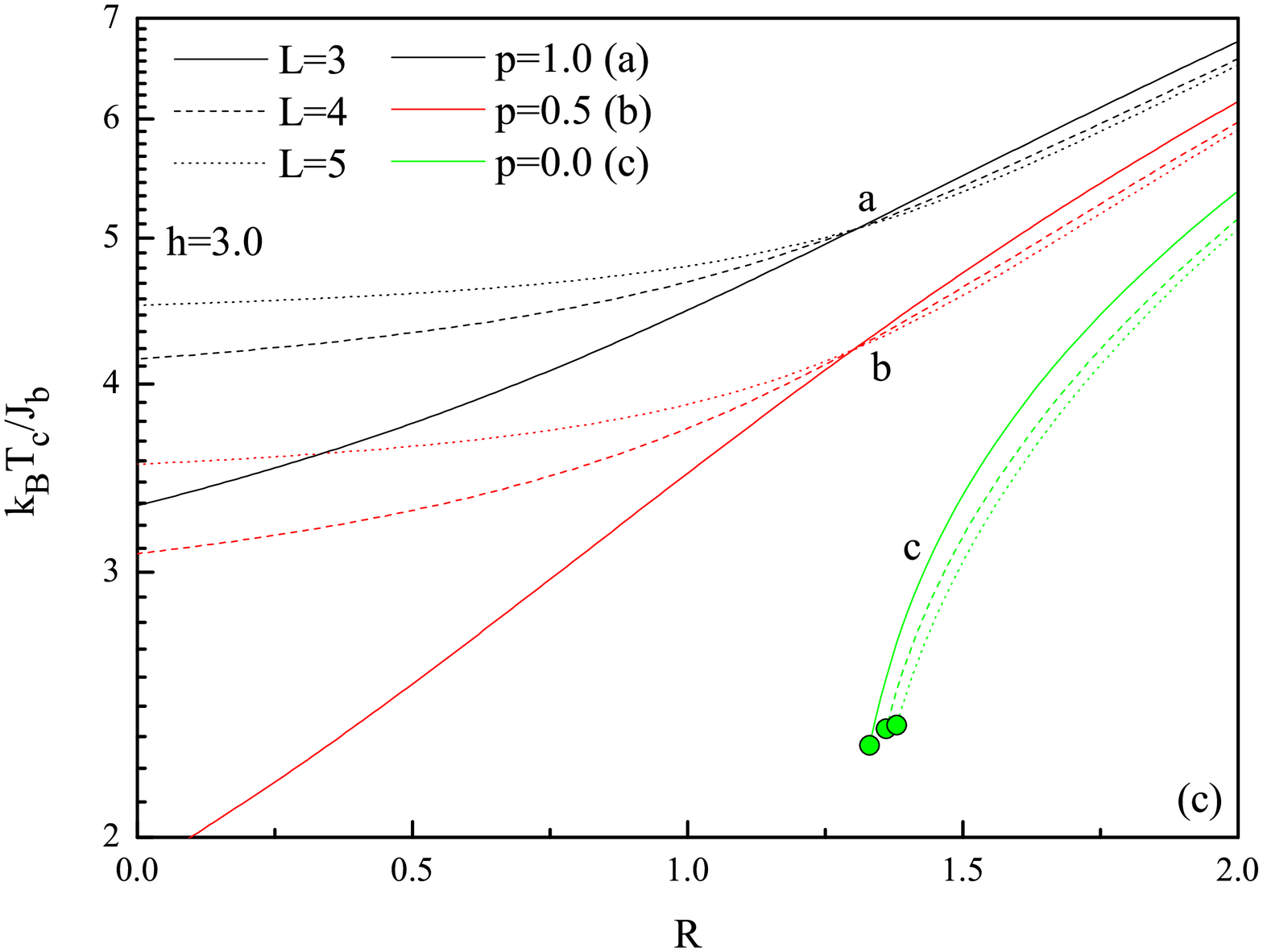}\\
\caption{Phase diagrams of the system in a $(k_{B}T_{c}/J_{b}-R)$
plane for various film thickness $L$ corresponding to trimodal
random field distribution with (a) $h=1.0$, (b) $2.0$, (c) $3.0$.
The letters on each set of curves denote the amount of disorder $p$.
Solid circles denote the tricritical points.} \label{fig4}
\end{figure}

Up to now, we have considered the presence of bimodal random fields.
In a trimodal random field distribution, in addition to the magnetic
field strength $h$, there is another parameter $p$ which controls
the amount of disorder acting on the system. According to Eq.
(\ref{eq2}), we have a pure system for $p=1.0$ whereas as $p$
decreases, then the effect of the bimodal random fields gradually
becomes dominant. In Fig. \ref{fig4}, phase diagrams in a
$(k_{B}T_{c}/J_{b}-R)$ plane are plotted for the films with
different thickness $L$ with a variety of $p$ values. By comparing
Figs. \ref{fig4}a-\ref{fig4}c, we see that the phase diagrams
corresponding to bimodal curves with $p=0$ evolve into pure system
with increasing $p$. When $p=0$, disorder is maximum, hence as $p$
increases, disorder effects are cleansed, tricritical points which
are observed in bimodal disorder configurations disappear, and the
transition temperature increases. However, the location of $R_{c}$
is insensitive to the presence of disorder whether the distribution
is bimodal or trimodal.

\begin{figure}[!h]
\center
\includegraphics[width=8.0cm]{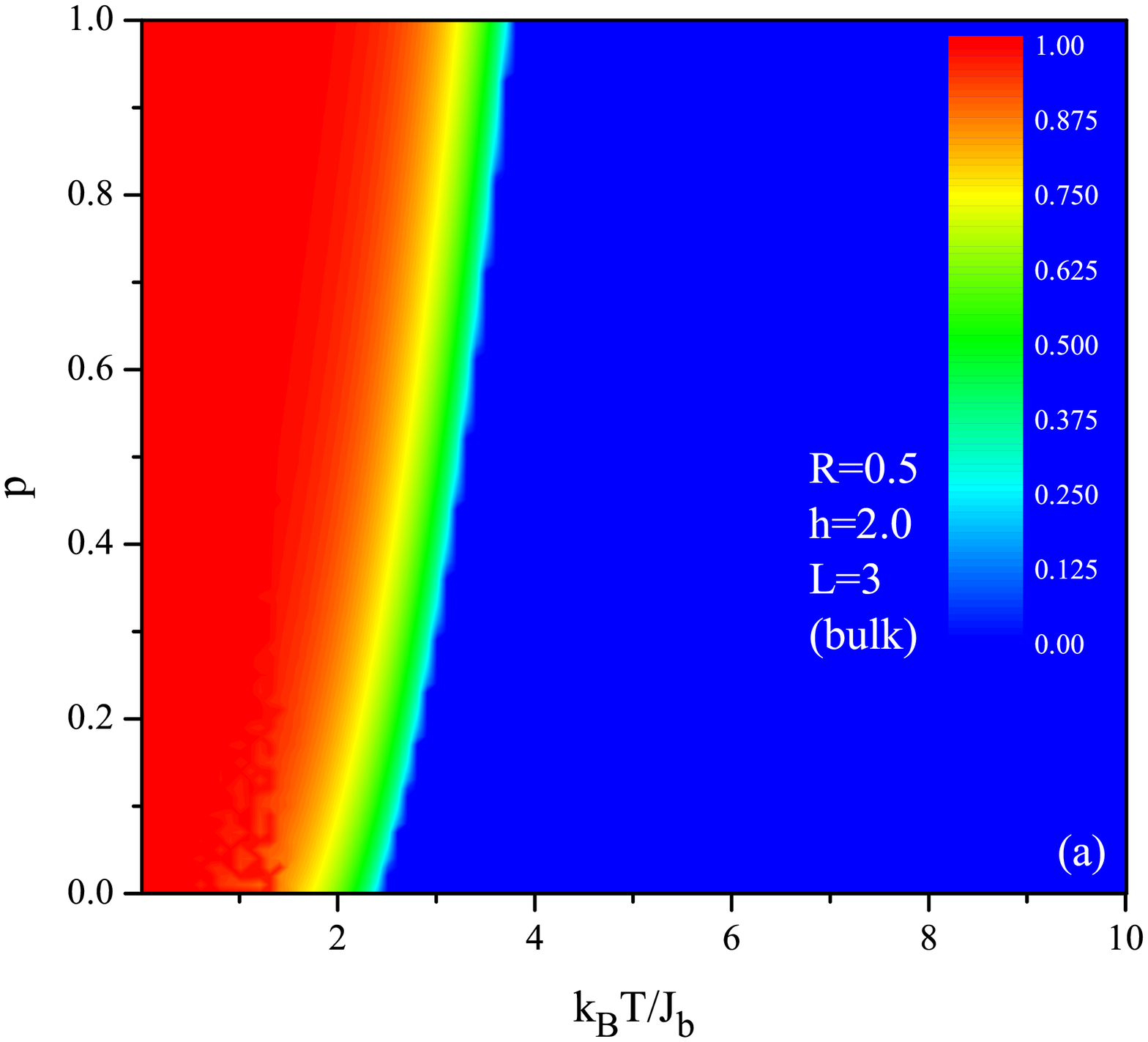}
\includegraphics[width=8.0cm]{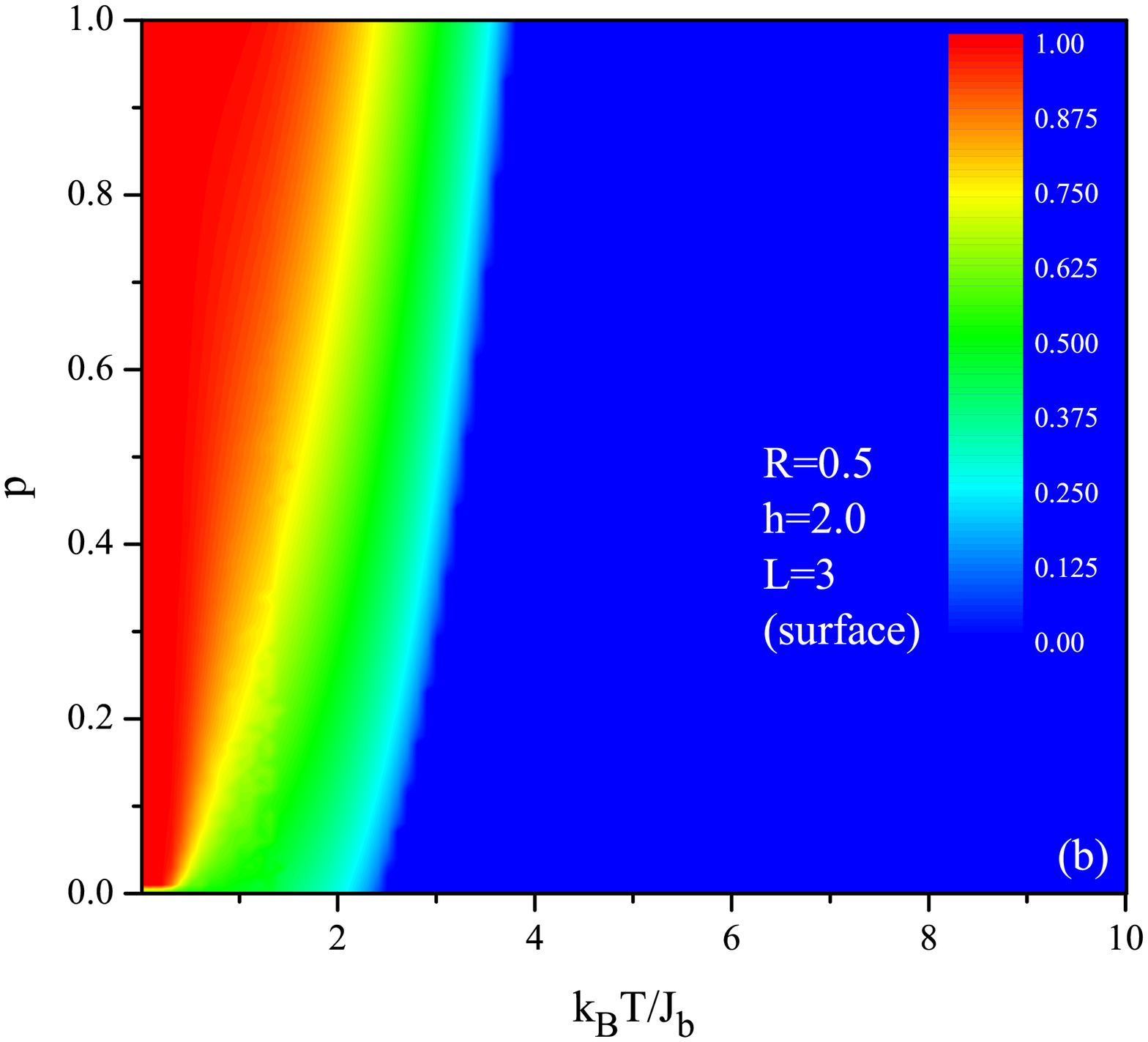}\\
\includegraphics[width=8.0cm]{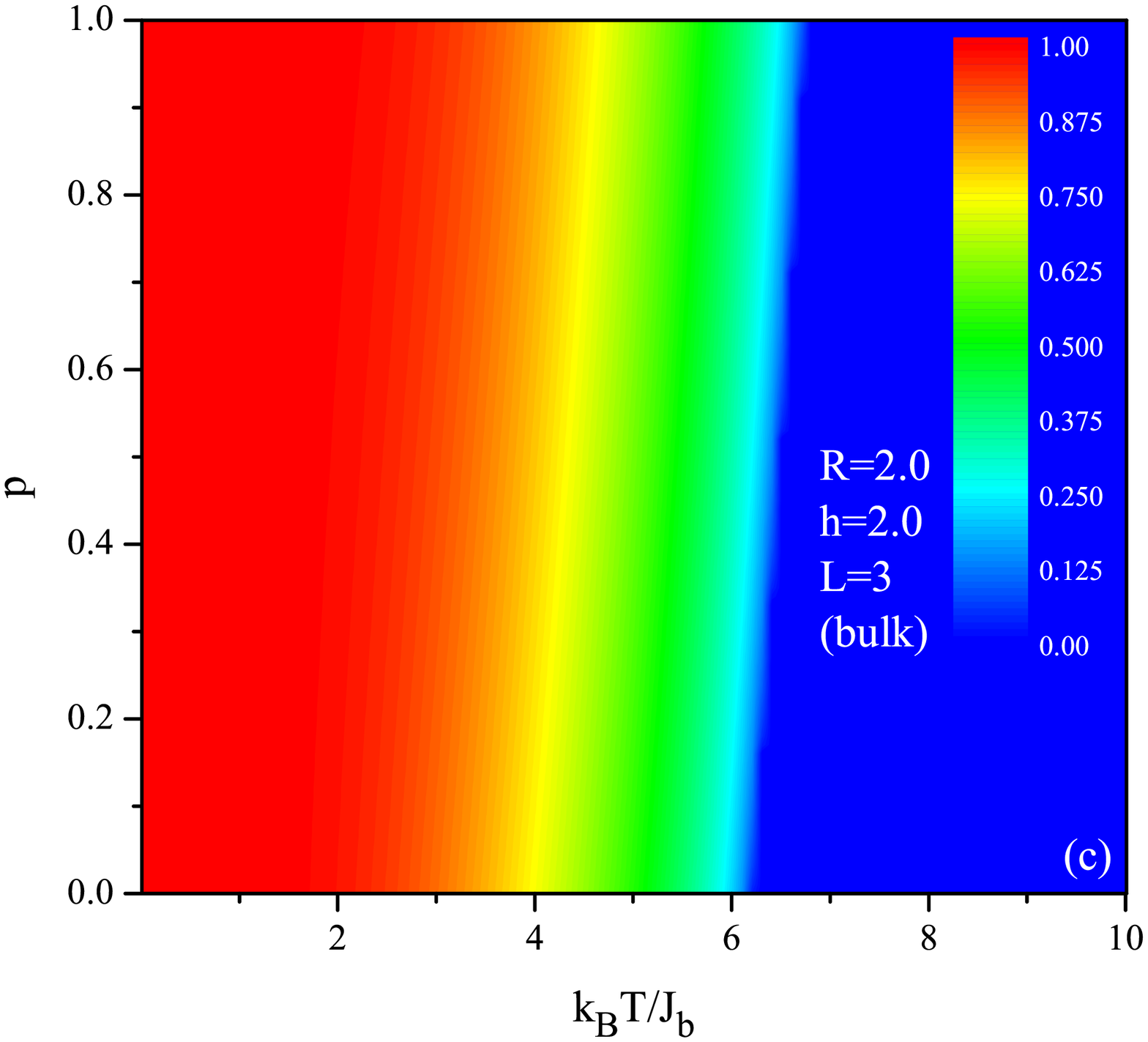}
\includegraphics[width=8.0cm]{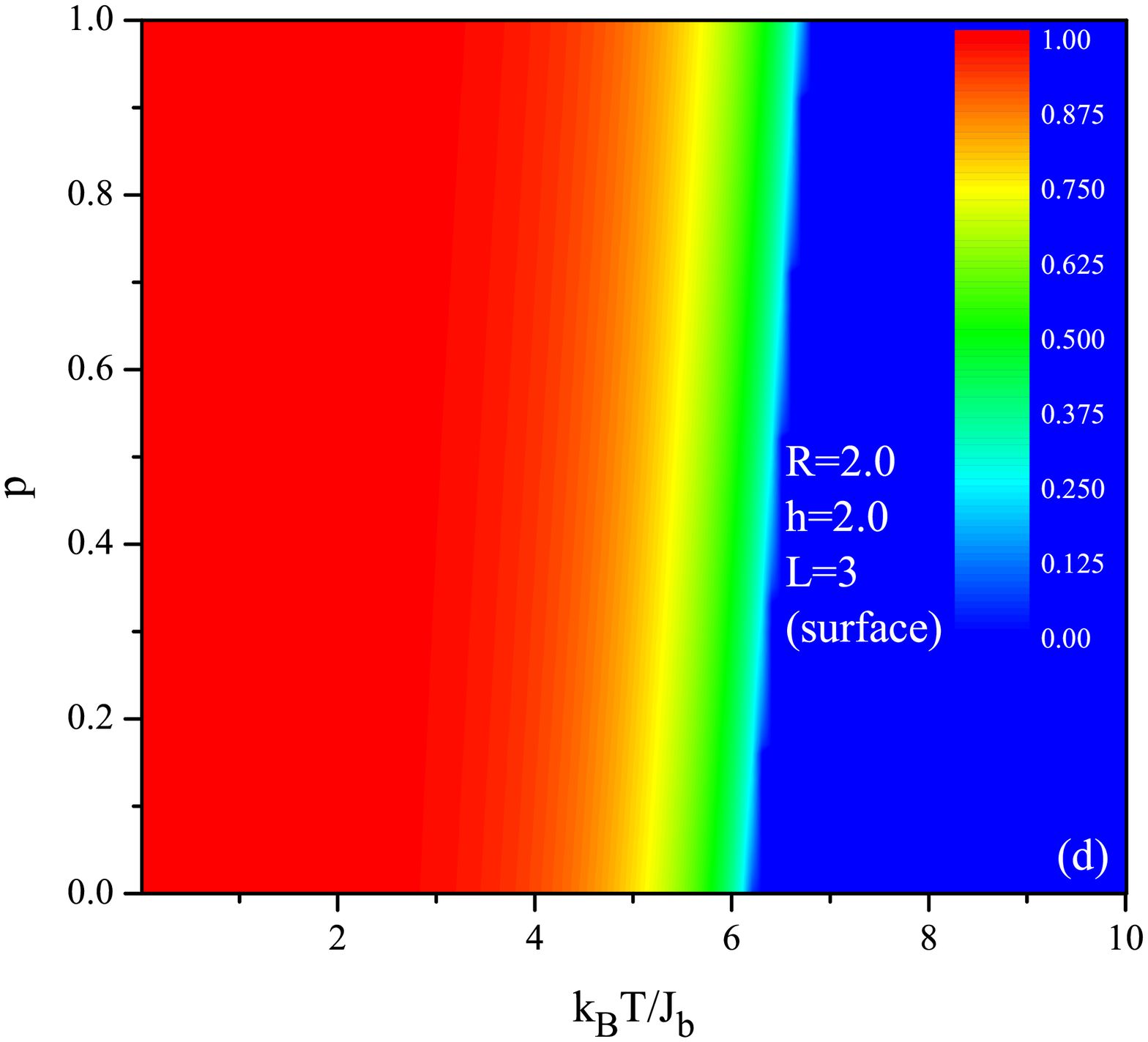}\\
\caption{Contour plot projections of bulk and surface magnetizations
in a $(p-k_{B}T/J_{b})$ plane with $L=3$ and $h=2.0$ in the presence
of (a), (b) reduced  $(R=0.5)$ and (c), (d) enhanced $(R=2.0)$
surface interactions.} \label{fig5}
\end{figure}

As a complementary investigation, bulk and surface magnetizations
with $L=3$ thick films in the presence of trimodal random fields
with $h=2.0$ as functions of disorder parameter $p$ and reduced
temperature $k_{B}T/J_{b}$ are shown in Fig. \ref{fig5}. As
expected, bulk (surface) region of the film is magnetically dominant
in ordinary (extraordinary) case. We can also conclude from this
figure that the transition temperature exhibits very slow variation
with disorder parameter $p$.

\begin{figure}[!h]
\center
\includegraphics[width=10.0cm]{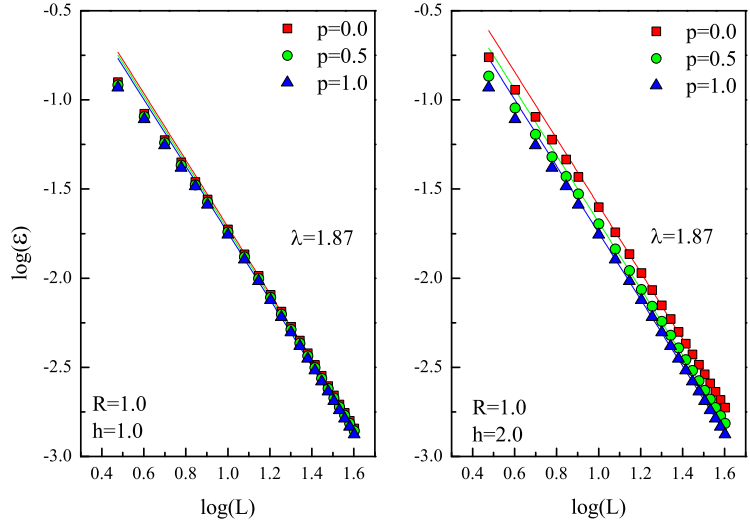}\\
\caption{Variation of the shift exponent $\lambda$ for moderate
surface coupling corresponding to some selected values of disorder
parameter $p$ with $h=1.0$ (left panel) and $h=2.0$ (right panel).}
\label{fig6}
\end{figure}

Finally, in Fig. \ref{fig6}, we examine the influence of trimodal
random fields on the shift exponent $\lambda$. One can easily
conclude from this figure that neither $p$, nor $h$ variations in
the system can affect the value of the shift exponent $\lambda$ (see
Table-\ref{table1}). Although the situation is depicted for moderate
values of reduced surface exchange interactions such as $R=1.0$, our
numerical data yield the same conclusion for any arbitrary $R<
R_{c}$ values.

\section{Conclusions}\label{conclude}
In conclusion, we have presented some results regarding the critical
behavior and universality properties of uniaxial ferromagnetic thin
films in the presence of bimodal and trimodal random fields. We have
given proper phase diagrams, magnetization profiles and variation of
the shift exponent $\lambda$ as a function of system parameters.

In our recent papers \cite{yuksel0,yuksel1}, we have found
that critical value of surface exchange coupling $R_{c}$ is a spin
dependent property, and variation of crystal field interactions, as
well as the presence of random crystal fields clearly affects the
value of $R_{c}$. However, the shift exponent $\lambda$ has been
found to be a spin "independent" property. Therefore, we have
concluded that a ferromagnetic spin-1/2 thin film is in the same
universality class with its spin-1 counterpart.

On the other hand, the main findings of the present work can be
summarized as follows:
\begin{itemize}
\item In the presence of bimodal random fields, the system exhibits
tricritical behavior for sufficiently large magnetic field
strengths. However, the linearized equations given in Eq.
(\ref{eq10}) are inapplicable for the first order transitions, and 
according to us, the tricritical behavior is a consequence of 
the nature of the bimodal random field distribution.
\item The location of $R_{c}$ at which the transition temperature
becomes independent of film thickness $L$ is insensitive to the presence of
disorder whether the distribution is bimodal or trimodal.
\item Regarding the universality properties, neither $p$, nor $h$ variations in
the system can affect the value of the shift exponent $\lambda$. In light of these findings, we can conclude that in the context of
the shift exponent, universality class of uniaxial ferromagnetic
thin films is independent of the presence of random magnetic fields.

\end{itemize}

We hope that the results presented in the present paper constitute
some preliminary ideas for future works based on more sophisticated
techniques.


\newpage

\newpage

\newpage

\newpage

\newpage

\newpage

\newpage

\begin{table}[!h]
\begin{center}
\begin{threeparttable}
\caption{The list of the extracted shift exponent $\lambda$ values
for several sets of random field parameters $h$ and $p$ in the
presence of weak $(R=0.5)$ and moderate $(R=1.0)$ surface
couplings. Corresponding residual error values due to fitting procedure are also listed in the rightmost column.}
\renewcommand{\arraystretch}{1.0}
\begin{tabular}{ccccc}
\thickhline
\ \ \ \ $R$ \ \ \ \  & \ \ \ \ $h$ \ \ \ \ & $p$  & \ \ \ \ $\lambda$ \ \ \ \  & \ \ \ \ error \ \ \ \ \\
\thickhline
0.5  & 0.0 (1.0, 2.0) & 0.0 (1.0)  & 2.01  & $2.0\times10^{-3}$ \\
0.5  & 1.0 & 0.0 & 2.01  & $9.72\times10^{-4}$ \\
0.5  & 2.0 & 0.0 & 2.01  & $6.23\times10^{-4}$ \\
1.0  & 0.0 & 0.0 & 1.87  & $6.0\times10^{-3}$ \\
1.0  & 1.0 & 0.0 & 1.87  & $4.7\times10^{-4}$ \\
1.0  & 2.0 & 0.0 & 1.87  & $6.0\times10^{-3}$ \\
1.0  & 1.0 & 0.5 & 1.87  & $4.47\times10^{-5}$ \\
1.0  & 2.0 & 0.5 & 1.87  & $5.95\times10^{-5}$ \\
\thickhline \\
\end{tabular}\label{table1}
\end{threeparttable}
\end{center}
\end{table}


\end{document}